\documentclass[aps,prb,reprint,superscriptaddress]{revtex4-2}

\usepackage{graphicx}
\usepackage{calc}
\usepackage{amsmath}
\usepackage{comment}
\usepackage{multirow}
\usepackage{xcolor}
\usepackage{float}
\usepackage[T1]{fontenc} 

\newcommand*{\sect}{Sec.\,}

\newcommand*{\fig}{Fig.\,}
\newcommand*{\eq}[1]{Eq.\,(#1)}
\newcommand*{\tab}{Tab.\,}
\newcommand*{\reftaken}{Ref.\,}
\newcommand*{\ie}{\emph{i.e.\,}}

\newcommand*{\degree}{$^{\circ}$}

\newcommand{\EFO}{ErFeO$_3$}
\newcommand{\JCNSMLZ}{Jülich Centre for Neutron Science at Heinz Maier-Leibnitz Zentrum, Forschungszentrum Jülich GmbH, Lichtenbergstraße 1, D-85747 Garching, Germany.}
\newcommand{\ISIS}{ISIS Facility, Rutherford Appleton Laboratory, Chilton, Didcot, OX11 0QX, United Kingdom}
\newcommand{\Aachen}{Institute of Crystallography, RWTH Aachen University, Jägerstraße 17–19, D-52066 Aachen, Germany}
\newcommand{\UJohannesburg}{Highly Correlated Matter Research Group, Physics Department, University of Johannesburg, P.O. Box 524, Auckland Park 2006, South Africa}

\begin{document}

\title{Magnetic excitation spectrum and hierarchy of magnetic interactions in ErFeO$_3$}

\author{Dnyaneshwar R. Bhosale}
\email{d.bhosale@fz-juelich.de}
\affiliation{\JCNSMLZ}

\author{Piotr Fabrykiewicz}
\affiliation{\JCNSMLZ}
\affiliation{\Aachen}

\author{Devashibhai~Adroja}
\affiliation{\ISIS}
\affiliation{\UJohannesburg}

\author{Martin Meven}
\affiliation{\JCNSMLZ}
\affiliation{\Aachen}

\author{Astrid Schneidewind}
\affiliation{\JCNSMLZ}

\author{Micha\l{} St\k{e}kiel}
\email{m.stekiel@fz-juelich.de}
\affiliation{\JCNSMLZ}

\date{\today}

\begin{abstract}
We report a comprehensive investigation of the excitation spectrum of \EFO ~orthoferrite by means of time-of-flight neutron spectroscopy. 
The spectrum consists of two distinct components: strongly dispersive spin wave excitations of the Fe$^{3+}$ sublattice spanning $\approx$~9 -- 65\,meV, and crystal electric field (CEF) excitations of Er$^{3+}$ ions below 36\,meV.
The observed spin wave dispersions and spectral weight are well captured within linear spin wave theory, enabling extraction of the key Fe~--~Fe exchange parameters.
Low-energy incident neutrons with their enhanced energy resolution, further revealed the dispersive character and splitting of Kramers-degenerate CEF levels.
We show that the dispersion is caused by the exchange coupling between Er$^{3+}$ ions, while the degeneracy is lifted by interactions between the Er$^{3+}$ and Fe$^{3+}$ sublattices.
We further explore the influence of dipolar and antisymmetric exchange interactions with the focus on the magnetic ground state of \EFO, with particular attention to the low-temperature spin arrangement.
Taken together, our results provide a detailed account of the spin dynamics in \EFO\ and reveal a hierarchy of interactions scales characteristic for orthoferrites.
\vspace{1em} \\
Physics Subject Headings (PhySH): Spin dynamics (Magnetism), Spin waves (Magnetism), Crystal field excitations (Structural properties), Time-of-flight neutron spectroscopy / Inelastic neutron scattering (Neutron techniques), Perovskites (Physical systems), Exchange interaction (Interaction and forces)

\end{abstract}


\maketitle
\newpage

\section{Introduction}


Rare-earth orthoferrites ($R$FeO$_3$ with $R$ = rare-earth element), and in particular \EFO, have garnered considerable attention due
to their rich magnetic phase diagrams and promising functionalities for next-generation spin-based technologies 
\cite{lin_evidence_2022, hortensius_coherent_2021, ma_low_2022, zhang_magnetic_2019, zic_coupled_2021, adachi_theory_2013, baltz_antiferromagnetic_2018, jungwirth_antiferromagnetic_2016, xu_observation_2022}. These materials exhibit a unique 
combination of high Néel temperatures (T$_\textrm{N}^\textrm{Fe}> 600$\,K), spin reorientation transitions, and strong 3d–4f magnetic interactions~\cite{kim_observation_2025, kumar_exchange_2025, Nikitin_Decoupled_spin_2018}, which 
give rise to tunable spin dynamics and complex magnetic structures \cite{li_terahertz_2023, zic_coupled_2021}. \EFO  ~is 
especially noteworthy for its capacity to host multiple magnetic phases, including temperature-driven reorientations of 
Fe$^{3+}$ spin configurations, and long-range ordering of Er$^{3+}$ moments at cryogenic temperatures. 
The weak ferromagnetic moment of canted Fe$^{3+}$ spins and induced Er$^{3+}$ moment compensates at the temperature around 45 K 
\cite{gorodetsky_magnetic_1973, deng_magnetic_2015, fita_temperature-driven_2022}. Such characteristics make \EFO  ~an ideal candidate for exploring 
phenomena like spin switching \cite{fita_temperature-driven_2022, ma_low_2022, zhang_magnetic_2019}, magnetoelectric coupling \cite{oh_anisotropic_2020}, multiferroicity \cite{oh_anisotropic_2020, deng_magnetic_2015, yokota_ferroelectricity_2015, juraschek_dynamical_2017}  and exchange bias \cite{fita_temperature-driven_2022} - key features for potential applications in magnonic
logic, memory devices, and antiferromagnetic spintronics~\cite{jungwirth_antiferromagnetic_2016}. Additionally,
intrinsic weak ferromagnetism stemming from spin canting render it a valuable model system for investigating the interplay between crystal and magnetic structure, magnetic anisotropy, and collective magnetic excitations. 
Thus, understanding the magnetic interactions that drive these properties in RFeO$_3$ orthoferrites is of fundamental interest as well as practical relevance for designing energy-efficient, high-speed spintronic components.


\EFO\ crystallizes in orthorhombically distorted perovskite structure with lattice parameters a=5.262\,\AA, b=5.583\,\AA, c=7.593\,\AA, and $Pbnm$ (non-standard setting of $Pnma$, \#62, space group) symmetry~\cite{marezio_crystal_1970}. The isolated Fe$^{3+}$ sublattice has higher, $Cmmm$, symmetry, but considering O$^{2-}$ ions forming distorted FeO$_6$ octahedra, and Er$^{3+}$ ions filling voids between them, reduces it to the $Pbnm$ space group. Neglecting both: (i)~a~6\% difference between $a$ and $b$ lattice parameters, and (ii)~a~1\% difference between the $c$ lattice parameter and $\sqrt{2}(a+b)/2$, one can describe the Fe$^{3+}$ sublattice using $Pm\bar{3}m$ space group, i.e. within non-distorted cubic perovskite approximation. While the $Cmmm$ symmetry leads to symmetry equivalent Fe$-$Fe bonds corresponding to exchange interactions: $J_1^\mathrm{ab}$, $J_1^\mathrm{c}$, $J_2^\mathrm{a}$, $J_2^\mathrm{b}$, and $J_2^\mathrm{d}$, the approximate $Pm\bar{3}m$ symmetry reduces those bonds/interactions into $J_1$ and $J_2$ as shown in \fig\ref{fig:Interactions}a.

Below T$_\mathrm{N}^\mathrm{Fe}$ = 620 K, the Fe$^{3+}$ ions' magnetic moments adopt a G-type antiferromagnetic ordering along the $a$-axis ~\cite{Wollan_Koehler} with a weak ferromagnetic component along the $c$-axis~\cite{koehler_neutron_1960}, or in short $\boldsymbol{G_x}F_z$ configuration \footnote{The notation follows Koehler~\cite{koehler_neutron_1960} and Bertaut~\cite{bertaut_repr_1968} works. 
The capitalized symbol denotes the type of ordering $F$ (ferromagnetic) $A/G/C$ (antiferromagnetic) and the subscript denotes the direction of the moment's component. The bold component is the one with the highest amplitude.}.
Over the temperature range $T^\mathrm{Fe}_\mathrm{SR}\approx$\,88 -- 98\,K, a spin reorientation transition (SRT)~\cite{ma_low_2022, belov_character_1976} occurs, in which the Fe$^{3+}$ moments rotate around $b$-axis to $F_x\boldsymbol{G_z}$ configuration, see Fig.~\ref{fig:Interactions}a. 
Below $T_\mathrm{N}^\mathrm{Er}$=4.1\,K the Er$^{3+}$ ions establish a $\mathcal{C}_z$ long-range antiferromagnetic ordering (Fig.~\ref{fig:Interactions}b) and further modify Fe$^{3+}$ ordering by inducing the component along the $b$-axis. 
Early studies report the $G_xG_y$~\cite{koehler_neutron_1960} and $F_xG_y\boldsymbol{G_z}$ configurations~\cite{gorodetsky_magnetic_1973}, but a more recent study~\cite{deng_magnetic_2015} determines the $F_xC_y\boldsymbol{G_z}$ configuration for magnetic structure of Fe$^{3+}$ below $T_\mathrm{N}^\mathrm{Er}$.


The Er$^{3+}$ fine structure in ErFeO$_3$ was studied by optical absorption in 1960s by Faulhaber \textit{et al.}~\cite{faulhaber_optical_1967} and Wood \textit{et al.}~\cite{Wood1969a, Wood1969b}.
Faulhaber \textit{et al.}~\cite{faulhaber_optical_1967} established the fine structure of the $^4$I$_{15/2}$ ground multiplet of Er$^{3+}$.
In the first approximation, it  consists of eight double degenerated levels, due to Kramers theorem.
However, a splitting of the ground-state of approximately 6.1~cm$^{-1}$ ($\sim$0.75~meV) was observed \cite{faulhaber_optical_1967}, affected by the onset of Er$^{3+}$ sublattice ordering below $T_\mathrm{N}^\mathrm{Er}$.
These results demonstrated that the ground-state splitting (6.44~cm$^{-1}$) arises from internal dipolar fields generated by Er$^{3+}$ (3.27~cm$^{-1}$) and Fe$^{3+}$ (1.39~cm$^{-1}$) sublattices along with Er-Fe exchange interactions (1.78~cm$^{-1}$).
Wood \textit{et al.}~\cite{Wood1969a} reported a change in the fine structure of the optical absorption spectrum near the spin-reorientation transition, with the Er$^{3+}$ level splittings decreasing by at least a factor of four upon crossing the reorientation region.
In a follow-up study~\cite{Wood1969b}, optical spectra were measured under an applied magnetic field, revealing a splitting of the Kramers doublets. Based on these observations, the authors concluded that the effective field at the Er$^{3+}$ site is dominated by the Er--Fe exchange interaction, while the magnetic dipolar contribution is comparatively small. 
Subsequent terahertz and ultrafast optical spectroscopy studies~\cite{Yamaguchi_2013_Terahertz, Mikhaylovskiy_2017_Selective} extended this understanding to the dynamic regime, revealing resonant magnetic and electric dipole-active modes in the frequency range 0.3--2\,THz (1.2--8.2\,meV) associated with low-energy excitations of Er$^{3+}$ transitions.
Furthermore, optical magnetospectroscopy studies on ErFeO$_3$ revealed a magnonic super-radiant phase transition in the terahertz and gigahertz frequency ranges at temperatures below 4\,K \cite{kim_observation_2025}. Their measurements considered the underlying magnetic interactions in the system, encompassing Fe–Fe, Er–Er, and Er–Fe exchange couplings, as well as magnetic anisotropies, within a mean-field Hamiltonian framework.


While optical studies in ErFeO$_3$ are limited to probing zone-center ($\Gamma$-point) excitations, inelastic neutron scattering (INS) allows to resolve magnetic excitations throughout the Brillouin zone, offering a more comprehensive understanding of the magnetic excitations.
Extensive INS investigations of $R$FeO$_3$ series ($R=$Y \cite{hahn_inelastic_2014}, Nd \cite{kumar_exchange_2025}, Tb \cite{ovsianikov_inelastic_2022}, Ho \cite{ovsyanikov_neutron_2020}, Tm \cite{skorobogatov_2020_low}, Yb \cite{nikitin_decoupled_2018}) have established some universal characteristics of the magnetic interactions in the family \cite{podlesnyak_RFO_2021}, driven by fundamental characteristics of transition-metal and rare-earth systems.
Fe sublattice is subject to relatively strong exchange interactions leading to coherent spin excitations, spinwaves, spanning the energy band up to 70\,meV, with the gap of 1-10\,meV induced by small easy axis anisotropy which changes across the spin reorientation.
These excitations are accurately modeled using a four-sublattice Hamiltonian incorporating nearest- and next-nearest-neighbor exchange interactions ($J_1$$\approx$4--5\,meV, $J_2$$\approx$0.1--0.3\,meV).
Fe$^{3+}$ ions are also subject to Dzyaloshinskii–Moriya interactions (DMI) due to strong spin-orbit coupling in the system, that accounts for spin canting and weak ferromagnetism and can introduce subtle features in the INS spectra \cite{hahn_inelastic_2014}, at the border of detectability limits.
In contrast, the rare-earth ions exhibit strong spin-orbit coupling resulting in CEF excitations~\cite{przenioslo_crystal_1995}, while weak $R$--$R$ interactions can result in a weakly dispersive character of low energy excitations \cite{podlesnyak_RFO_2021}.
In particular, NdFeO$_3$ \cite{kumar_exchange_2025} shows dispersive character of the 0.5\,meV excitation along main crystallographic directions modeled by dominating in-plane Nd--Nd interactions. In TmFeO$_3$ \cite{skorobogatov_2020_low} two lowest excitations below 8\,meV exhibit pronounced dispersion along both directions of the (0\textit{KL}) plane, while YbFeO$_3$ \cite{Nikitin_Decoupled_spin_2018} shows weakly dispersive dispersion confined to the (00\textit{L}) direction.
For the latter the spin-chain model is established for the  Yb$^{3+}$ subsystem, which evolves from a field-polarized magnon regime below the spin-reorientation temperature of 7.6~K to a spinon-like continuum above it, reflecting the strong influence of Fe$^{3+}$ molecular fields on rare-earth anisotropy.

In ErFeO$_3$, previous neutron scattering studies reported only the low-energy dynamics and revealed temperature-dependent shifts and splitting of low-energy excitations~\cite{zic_coupled_2021, shapiro_neutron-scattering_1974}.
In particular, the 6\,meV excitation is split by 0.5\,meV and shows mode-dependent polarization while the lowest CEF level shifts from 0.4\,meV at 6\,K to 0.8\,meV at 1.5\,K, reflecting the sensitivity of the Er$^{3+}$ sublevel structure to the evolving magnetic environment shaped by both Er$^{3+}$ and Fe$^{3+}$ magnetic orders \cite{zic_coupled_2021}. 
Further, Shapiro \textit{et. al} \cite{shapiro_neutron-scattering_1974} studied the temperature-dependent low-energy spin dynamics across the SRT region and treated the excitations observed across 0 -- 5 meV as a 'low energy magnons' and discussed the softening of the observed magnon mode as a function of temperature~\cite{shapiro_neutron-scattering_1974}.

These observations underscore the need for a comprehensive investigation of the magnetic excitation spectrum in ErFeO$_3$, extending beyond the previously explored low-energy regime \cite{zic_coupled_2021, obrien_giant_2023} dominated by Er$^{3+}$ crystal-field dynamics to include the high-energy spin wave excitations arising from the Fe$^{3+}$ sublattice. In this work, we reveal the magnetic excitation spectrum of ErFeO$_3$ across the relevant energy scales and establish a unified model of magnetic interactions encompassing Fe–Fe, Er–Er, and Er–Fe couplings, consistent with frameworks reported for other orthoferrites. Importantly, we highlight the role of dipolar along with exchange interactions as an effective Er–Fe coupling mechanism, driven by the large magnetic moment of Er$^{3+}$, which influences the Fe$^{3+}$ spin subsystem and the resulting magnetic ground state. By capturing these intertwined magnetic interactions, our study establishes a hierarchy of energy scales governing the magnetic dynamics in ErFeO$_3$, bridging low- and high-energy excitations and laying the foundation for the design of orthoferrite-based materials optimized for magnonic and spin-transport applications.

\section{Experimental Details}

The  single-crystal \EFO\ used in this study was grown by the floating zone method from a polycrystalline ingot, using the protocol used for synthesis of various $R$FeO$_3$ orthoferrites described in \reftaken\cite{ovsianikov_PhD_2023}.

The cylindrical sample (7\,mm length, 6\,mm diameter) was oriented in the ($H0L$) scattering (horizontal) plane and mounted in an orange-type He cryostat.
The sample was measured at 1.5 and 6\,K, spanning below and above the N\'{e}el temperature of Er$^{3+}$, as well as at 120 K, above the spin reorientation transition of Fe$^{3+}$.

Inelastic neutron scattering measurements were carried out at the MERLIN thermal neutron time-of-flight spectrometer at the ISIS Neutron and Muon Source, UK \cite{bewley_merlin_2006}.
The MERLIN spectrometer was operated with the Gd chopper package, enabling repetition-rate multiplication (RRM), which allows for simultaneous measurements with multiple neutron incoming energies ($E_i$).
To probe the broad energy range relevant to both spin wave and CEF excitations, measurements were performed at chopper frequencies of 300\,Hz (at 1.5, 6, and
120\,K) and 350\,Hz (at 6\,K).
At 300\,Hz, the available incident energies $E_i$ were 100, 33.44, 16.54, and 9.84\,meV; while at 350~Hz, $E_i$ energies included 160, 49.89, 24.01, and 
14.07\,meV.
The sample was rotated in steps of 1\degree~over the range of 0 -- 90\degree ~to determine the spectral weight as a function of momentum transfer vector and energy
[$S(\vec{Q}, E)$].
Momentum averaging was performed to determine $S(Q, E)=\big<S(\vec{Q}, E)\big>_{|\vec{Q}|=Q}$ and investigate non-dispersive excitations.

Data reduction was performed in Mantid~\cite{Mantid_url, Mantid_paper} and Horace~\cite{Horace_paper}, accessed via the ISIS Data Analysis as a Service \cite{Data_server_ISIS}.
Data analysis based on the linear spinwave theory \cite{SpinW_paper} utilized our own python implementation \cite{stekiel_spinwaves_2025} while calculations involving dipolar excitations were done in \textsc{Sunny} \cite{sunny_2025}, crystal-field excitations were investigated with \textsc{PyCrystalField}~\cite{PyCrystalField_paper}.
\textsc{VESTA}~\cite{VESTA_paper} was used for crystal structure visualization.

\section{Results}

\begin{figure}
    \includegraphics[width=\columnwidth]{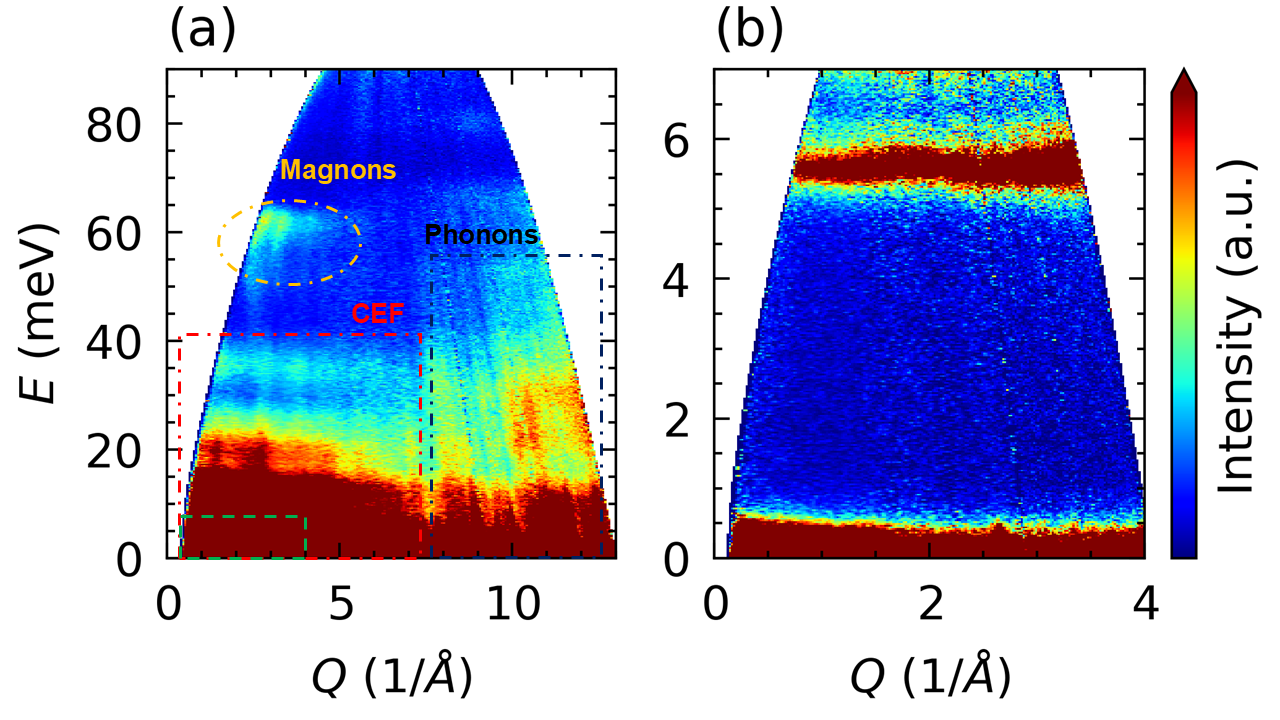}
    \caption{
    Overview of powder-averaged inelastic neutron scattering (INS) spectrum of ErFeO$_3$ collected with an incident energy $E_{i} = 100$\,meV (a) and $E_{i} = 9.84$\,meV (b) at $T = 6 $\,K. Panel (a) shows dominated regions for (i) dispersive spin wave excitations (magnons) from the Fe$^{3+}$ sublattice at low momentum transfer $Q$, (ii) almost non-dispersive crystal electric field (CEF) excitations of Er$^{3+}$ ions, and (iii) phonon contributions at higher $Q$. Panel (b) reveals further details of CEF excitations with weak dispersive character. The rectangle in the bottom left corner of panel (a) shows the region covered by panel (b).
    }
    \label{fig:Powder-average}
\end{figure}

The representative powder-averaged inelastic neutron scattering spectrum of ErFeO$_3$, measured with an incident energy of $E_i = 100$\,meV over a wide momentum-transfer range, is shown in Fig.~\ref{fig:Powder-average}a. 
The spectrum reveals three distinct components: (i) strongly dispersive spin wave excitations originating from the Fe$^{3+}$ sublattice with a band maximum near 62\,meV, (ii) crystal electric field (CEF) excitations associated with Er$^{3+}$ ions below 36~meV, and (iii) phonon contributions corresponding to lattice vibrations appearing at higher momentum transfers ($Q > 8$~\AA$^{-1}$).
The low energy spectrum ($E_i = 9.84$\,meV) due to its enhanced resolution further revealed weak dispersive character of CEF excitations, as shown in Fig.~\ref{fig:Powder-average}b. 
Focusing on the magnetic interactions that govern excitations in ErFeO$_3$, the system can be described by a Hamiltonian comprising three primary contributions: interactions within the Fe$^{3+}$ sublattice ($\mathcal{H}_\mathrm{Fe}$), within the Er$^{3+}$ sublattice ($\mathcal{H}_\mathrm{Er}$), and between the two sublattices ($\mathcal{H}_\mathrm{Fe-Er}$):

\begin{equation} \label{eq:hamiltonian_general}
\begin{split}
\mathcal{H} = & \overbrace{ \mathcal{H}_\mathrm{Fe-Fe}^\mathrm{Ex} + \mathcal{H}_\mathrm{Fe}^\mathrm{SIA} + \mathcal{H}_\mathrm{Fe-Fe}^\mathrm{DMI}}^{\mathcal{H}_\mathrm{Fe}} \\
              & + \underbrace{\mathcal{H}_\mathrm{Er}^{CEF} + \mathcal{H}_\mathrm{Er-Er}^{Ex}}_{\mathcal{H}_\mathrm{Er}}+ \underbrace{\mathcal{H}_\mathrm{Fe-Er}^\mathrm{Ex} + \mathcal{H}_\mathrm{Fe-Er}^\mathrm{DIP}}_{\mathcal{H}},
\end{split}
\end{equation}
where Ex stands for exchange interactions, SIA for single-ion anisotropy and DIP for dipolar interactions.

We present the experimental observations together with their corresponding modelling, organised according to the decomposition of the full Hamiltonian in Eq.~\ref{eq:hamiltonian_general}.
Table~\ref{tab:hamiltonian} summarises all the terms that lead to the full Hamiltonian together with refined values of interaction parameters.
The high-energy, strongly dispersive spin wave excitations originating from
the Fe$^{3+}$ sublattice, their temperature evolution, and the associated energy gap are discussed in Sec.~\ref{sect:spin waves}. 
The medium-energy CEF excitations of the Er$^{3+}$ ions, including their weakly dispersive nature and splitting of Kramers doublets, are analyzed in Sec.~\ref{sect:CEF}. Finally, the final parts of Sec.~\ref{sect:CEF} address the inter-sublattice interactions between Fe$^{3+}$ and Er$^{3+}$ ions.

\begin{figure}
    \includegraphics[width=\columnwidth]{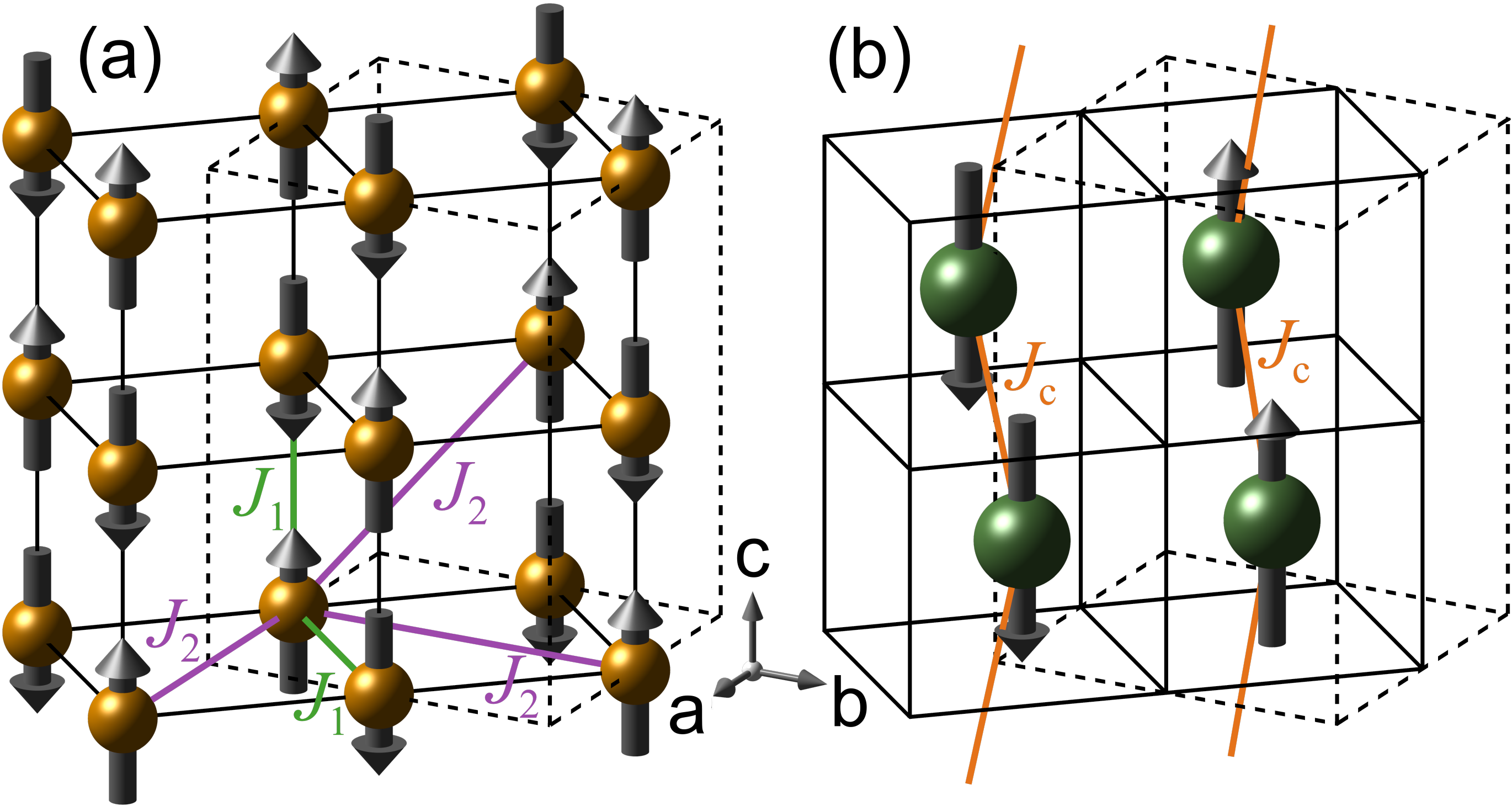}
    \caption{
    Overview of magnetic exchange interactions in ErFeO$_3$.
    (a)~Magnetic unit cell (dotted line) of ErFeO$_3$ below spin reorientation transition (SRT), highlighting the exchange interactions within Fe$^{3+}$ sublattice (4$b$ Wyckoff position with $\bar{1}$ site symmetry point group). Arrows represent the main $G_z$ ($\Gamma_2$, $Pbn'm'$) component of Fe$^{3+}$ magnetic moments, characterised by aniferromagnetic arrangement of all nearest neighbours. Non-equivalent interactions up to 5$^\textrm{th}$ coordination shell are shown. Those are aggregated into nearest-neighbor ($J_1$) and next-nearest-neighbor ($J_2$) interactions, arising from two district Fe--Fe distances in approximate cubic perovskite symmetry, which unit cells are given with the solid line.
    (b)~Magnetic unit cell of ErFeO$_3$ below $T_\mathrm{N}^\mathrm{Er}$, highlighting the exchange interactions within Er$^{3+}$ sublattice (4$c$ Wyckoff position with $..m$ site symmetry point group). Arrows represent the $\mathcal{C}_z$ ($\Gamma_1$, $Pbnm.1$) ordering of Er$^{3+}$ magnetic moments, characterized by ferromagnetic arrangement of Er$^{3+}$ moments within chains along the $c$-axis, and aniferromagnetic arrangement of nearest neighbours chains. Nearest neighbour intrachain exchange interactions are marked with $\mathcal{J}_c$. Weak interchain interactions are not shown as they are not required to describe the INS spectra.
    }
    \label{fig:Interactions}
\end{figure}

\subsection{Iron sublattice excitations}
\label{sect:spin waves}

\begin{figure}
    \includegraphics[width=\columnwidth]{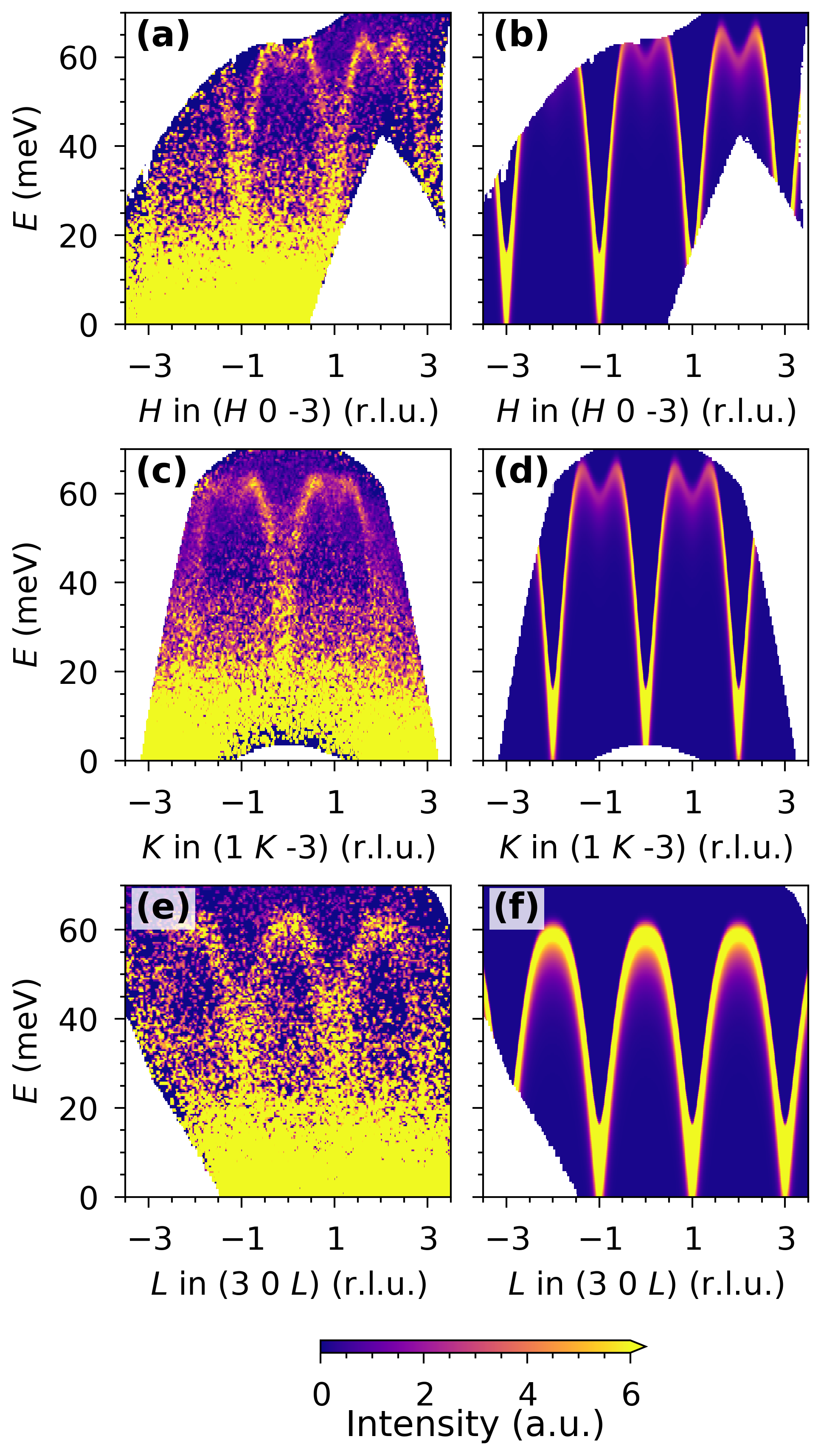} 
    \caption{Spin wave dispersion relations of ErFeO$_3$ at 120\,K. Experimental data (a,~c,~e) at 120\,K and model (b,~d,~f) are displayed along the high-symmetry $Q$-directions (a,~b) (\textit{H}~0~–3), (c,~d) (1~\textit{K}~–3), and (e,~f) (3~0~\textit{L}). The data reveal pronounced variations in dispersion curves and spectral weight across different directions in reciprocal space and are used to extract the exchange interactions in the Fe-sublattice and model the dispersions according to the $\boldsymbol{G_x}F_z$ spin arrangement. Intensity cut-off was used to visualise spinwave dispersions in panels (a,~c,~e), and as a result, the low energy range of energy dominated by CEF excitation is oversaturated. The experimental maps were obtained by integrating the measured signal in $\pm$0.05\,r.l.u. range in directions perpendicular to the selected cut.}
    \label{fig:magnons_120K}
\end{figure}

The spin wave band associated with the Fe$^{3+}$ sublattice extends from approximately 9 to 62~meV, with a single, well-defined excitation branch observed across the entire energy range. 
The overall dispersion remains unchanged within experimental resolution across all measured temperatures, as exemplified by the representative spectra collected at 120~K shown in Fig.~\ref{fig:magnons_120K}.
The persistence of the dispersion shape at all measured temperatures indicates that the underlying exchange interactions governing the Fe$^{3+}$ spin waves remain unchanged across the magnetic transitions in ErFeO$_3$. 
The apparently single excitation branch is, in fact, doubly degenerate, given the antiferromagnetic arrangement of the Fe sublattice.
There is also a doubling of the spinwave periodicity relative to the crystallographic lattice, reflecting Brillouin zone unfolding typical for antiferromagnetic spin waves.
Such behavior suggests an apparent higher $Cmmm$ symmetry of the Fe sublattice and a dominant $G$-type component of the magnetic order. 
The presence of DMI, manifested through a small canting of the Fe moments that restores the $Pbnm$ symmetry, was shown in earlier rare-earth orthoferrites studies to induce weak spectral intensity in the otherwise silent high-energy branch~\cite{hahn_inelastic_2014} and to cause fine structure within the spin wave gap~\cite{park_low-energy_2018}.
While these subtle DMI-related features remain below the detection limit of present measurements, their influence on the magnetic ground state will be discussed in detail in Sec.~\ref{sect:GS}. 

\begin{figure}
    \includegraphics[width=\columnwidth]{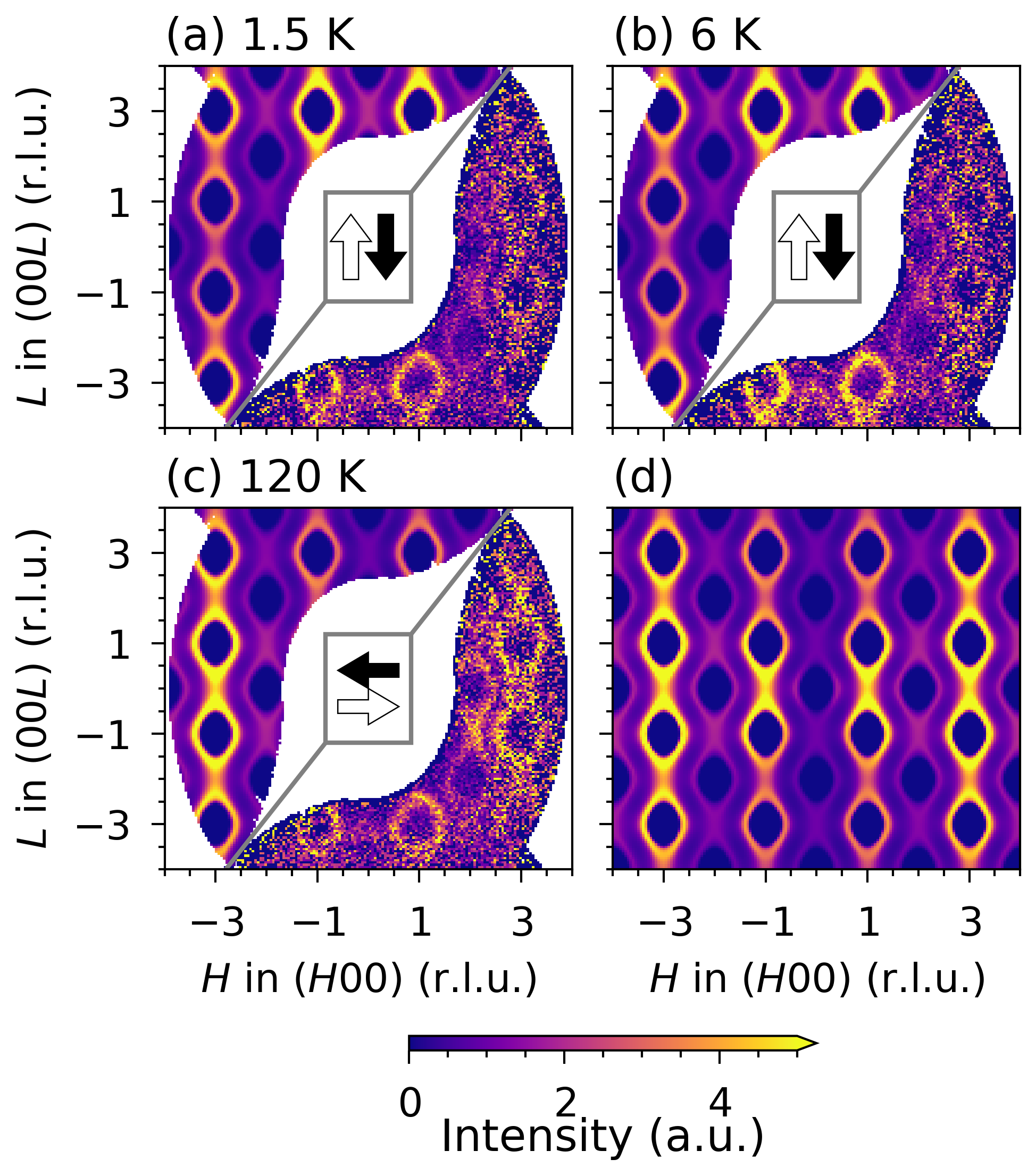} 
    \caption{
    High-energy spectral weight of \EFO. Panels show intensity integrated over the energy range 50–62\,meV, as a function of momentum transfer with (\textit{H}0\textit{L}) coordinates. Panels (a-c) show measured (lower-right) and modeled (upper-left) spectral weight at 1.5\,K (a), 6\,K (b), and 120\,K (c), as well as schematic spin arrangement of magnetic moments in the central inset. To facilitate comparison between the calculated and measured intensity the modeled intensity is constrained to the momentum transfers covered by measurements. Panel (d) shows unconstrained intensity of the 120\,K model as in panel (c).
    Reconstructions highlight temperature-dependent variations in the spectral weight arising from the change in magnetic structure as indicated by arrows, while the dispersion relations are unaffected.
    }
    \label{fig:magnons_Ecut} 
\end{figure}

While the overall dispersion relations remain unchanged, the spectral weight distribution exhibits a pronounced temperature dependence; distinct between 120\,K and 6\,K, yet nearly identical at 6\,K and 1.5\,K.
This effect is clearly visualized in the constant-energy maps presented in Fig.~\ref{fig:magnons_Ecut}, which capture the spectral weight redistribution across temperatures.
The variation arises from the SRT occurring between 88 and 98\,K, during which the dominant antiferromagnetic $G$-type component of the Fe$^{3+}$ magnetic order rotates from $G_x$ above 98\,K to $G_z$ below 88\,K~\cite{ma_low_2022}. 
This reorientation redistributes the spectral weight of the spin waves, clearly seen in constant-energy maps in Fig.~\ref{fig:magnons_Ecut}b (6\,K) and Fig.~\ref{fig:magnons_Ecut}c (120\,K). 
The onset of the long-range order of Er sublattice below T$_N^{Er}$ = 4.1\,K does not modify the spectrum significantly, as compared in Fig.~\ref{fig:magnons_Ecut}a (1.5\,K) and  Fig.~\ref{fig:magnons_Ecut}b (6\,K).
The temperature-dependent redistribution of spectral weight is well reproduced by our linear spin wave theory simulations, shown in the upper-left parts of Fig.~\ref{fig:magnons_Ecut}a–c, constrained to the experimentally measured momentum-transfer range. 
For completeness, \fig\ref{fig:magnons_Ecut}d displays the unconstrained intensity from panel \fig\ref{fig:magnons_Ecut}c.

\begin{figure}
    \includegraphics[width=\columnwidth]{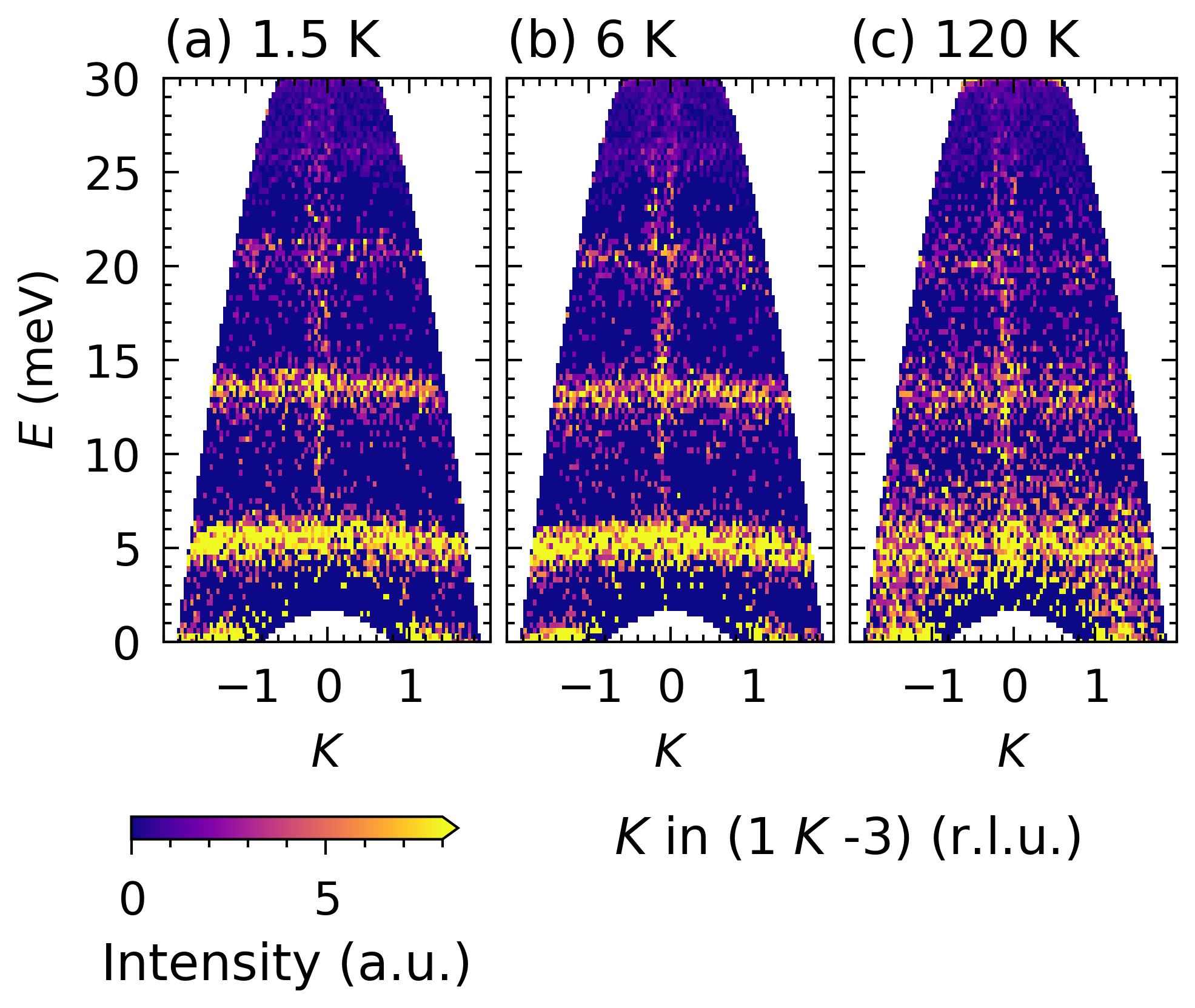}
    \caption{
    Intermediate energy magnetic excitations in \EFO\ measured at (a) 1.5\,K, (b) 6\,K, and (c) 120\,K. At each temperature spectra consist of strongly dispersive Fe$^{3+}$ spin wave excitation centered around K=0 revealing a pronounced energy gap of $\approx$9 meV, and non-dispersive Er$^{3+}$ crystal electric filed (CEF) excitations. There is little difference between spectra measured at 1.5 and 6\,K (a,b) as the principal component of the Fe$^{3+}$ moments is the same. The spectrum at 120\,K (c) features broadened excitations as compared to
    lower temperatures. The maps were obtained by integrating the measured signal in $\pm$0.05\,r.l.u. range in $H$ and $L$ directions.
    }
    \label{fig:magnons_gap}
\end{figure}

Measurements conducted with a lower incident neutron energy of E$_i$ = 33.4\,meV reveal a spin wave energy gap of approximately 9~meV, as shown in Fig.~\ref{fig:magnons_gap}, along with distinct low-lying CEF excitations. 
Within the experimental energy resolution of $\delta E$ = 1.5\,meV at zero energy transfer, the energy gap remains unchanged between 1.5~K and 6~K, consistent with earlier observations~\cite{zic_coupled_2021}.
Notably, the gap persists at roughly 9\,meV even at 120\,K, indicating that the spin reorientation transition has little influence on the anisotropy-driven energy gap.
Nevertheless, higher-resolution measurements would be required to establish the accuracy in spin wave gap energy measurement.

To model the observed spin wave excitations we employed a minimal $J_1$–$J_2$ Heisenberg model $\mathcal{H}_\mathrm{Fe}^\mathrm{Ex} + \mathcal{H}_\mathrm{Fe}^\mathrm{SIA}$ previously reported for various orthoferrites \cite{hahn_inelastic_2014, shapiro_neutron-scattering_1974,GUKASOV_Neutron_1997, White_Light_1982, Nikitin_Decoupled_spin_2018}, which accounts for nearest- (NN, $J_1$) and next-nearest-neighbor (NNN, $J_2$) Fe--Fe exchange interactions (Fig.~\ref{fig:Interactions}a) as well as single-ion anisotropies ($K_x$, $K_z$):
\begin{equation}
    \label{eq:H_Fe-Fe_Ex}
    \mathcal{H}_\mathrm{Fe}^\mathrm{Ex} = J_1 \sum_{i,j \in \mathrm{NN}} \mathbf{S}_i \cdot \mathbf{S}_j + J_2 \sum_{i,j \in \mathrm{NNN}} \mathbf{S}_i \cdot \mathbf{S}_j,
\end{equation}
\begin{equation}
    \label{eq:H_Fe_SIA}
    \mathcal{H}_\mathrm{Fe}^\mathrm{SIA} = K_x \sum_i (\mathbf{S}_i^x)^2 + K_z \sum_i (\mathbf{S}_i^z)^2,
\end{equation}

\noindent where $\mathbf{S}$ are Fe$^{3+}$ magnetic moments operators, and $\mathbf{S}^x$ ($\mathbf{S}^z$) are their $x$ ($z$) components. 
This model was implemented within the framework of linear spin wave theory (LSWT) \cite{toth_linear_2015} and found to reproduce the experimentally observed dispersion relations (curvature, bandwidth, spectral weight, and energy gap) with remarkable accuracy.
The exchange parameters obtained from the refinement of the model to the observed magnon energies were $J_1=5.09\,(46)$\,meV and $J_2=0.31\,(18)$\,meV.
The gap in the observed spectra suggests easy axis anisotropy, that is negative anisotropy value $K<0$.
To account for the spin reorientation, we use an effective model, where above the spin reorientation we use $K_x$=$K$ and $K_z$=0 to reproduce the arrangement of Fe moments along the $a$-axis, while below the spin reorientation $K_x$=0 and $K_z$=$K$ where moments align with the $c$-axis.
We obtained $K=-0.117\,(35)$\,meV, which gives rise to the 9\,meV energy gap.
This approach reproduces the spinwave dispersions very well (\fig\ref{fig:magnons_120K}), including the redistribution of the spectral weight due to the spin reorientation transition (\fig\ref{fig:magnons_Ecut}).

\subsection{Erbium sublattice excitations}
\label{sect:CEF}

\begin{figure}
    \includegraphics[width=\columnwidth]{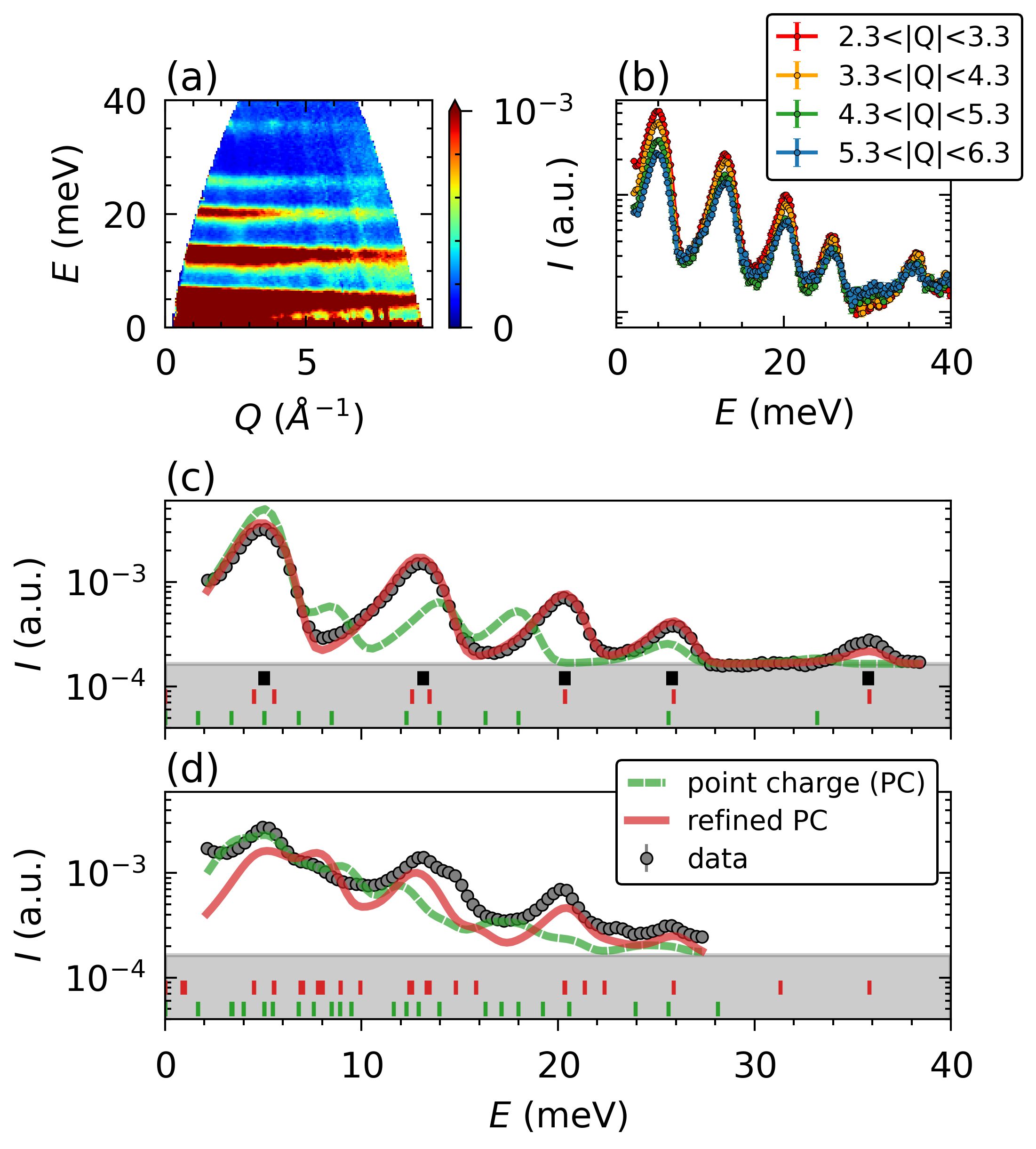}
    \caption{Intermediate energy spectra of ErFeO$_3$ resolving the CEF excitations. (a) Inelastic neutron scattering spectrum as a function of direction-averaged momentum transfer $Q$, recorded with an incident neutron energy of 49.9 meV. The spectrum reveals multiple excitations with (b) corresponding dynamical structure factor $S(Q, \omega)$ presented as constant-$Q$ cuts, integrated over selected $Q$ ranges for an energy transfer window of 2 – 40 meV. The spectral weight of these non-dispersive bands diminishes with increasing $Q$, characteristic for CEF excitations.
    (c – d) Fits to the spectra measured at 6\,K (c) and 120\,K (d) and integrated over $Q$<8\,\AA$^{-1}$ range. Fits include the initial (green, dotted line) and refined spectra (red, solid), where the initial spectrum is based on the point-charge model.
    Note that at 120\,K the spectrum was measured with E$_i$ of 33.4 meV.
    \label{fig:CEF-powder}}
\end{figure}

To investigate CEF excitations in ErFeO$_3$ we explored momentum-averaged excitation spectra up to 140\,meV and found five CEF excitations, all below 40\,meV, see~\fig\ref{fig:CEF-powder}.
Measurements at 6\,K performed with an incoming energy of 49.89\,meV and resolution of approximately $2$\,meV  (at zero energy transfer) reveal excitation energies 5.8, 14.0, 20.3, 25.7, and 35.9\,meV, consistent with previous reports \cite{Wood1969a, faulhaber_optical_1967, zic_coupled_2021, obrien_giant_2023}.
These excitations exhibit a typical magnetic character, with intensities diminishing with increasing momentum transfer $Q$ due to the magnetic form-factor, see Fig.~\ref{fig:CEF-powder}b with constant-$Q$ cuts of integrated intensity displayed on the logarithmic scale.
They are well separated from phonon contributions (\fig\ref{fig:Powder-average}a), which dominate the high-$Q$ region.

\begin{figure}
    \includegraphics[width=\columnwidth]{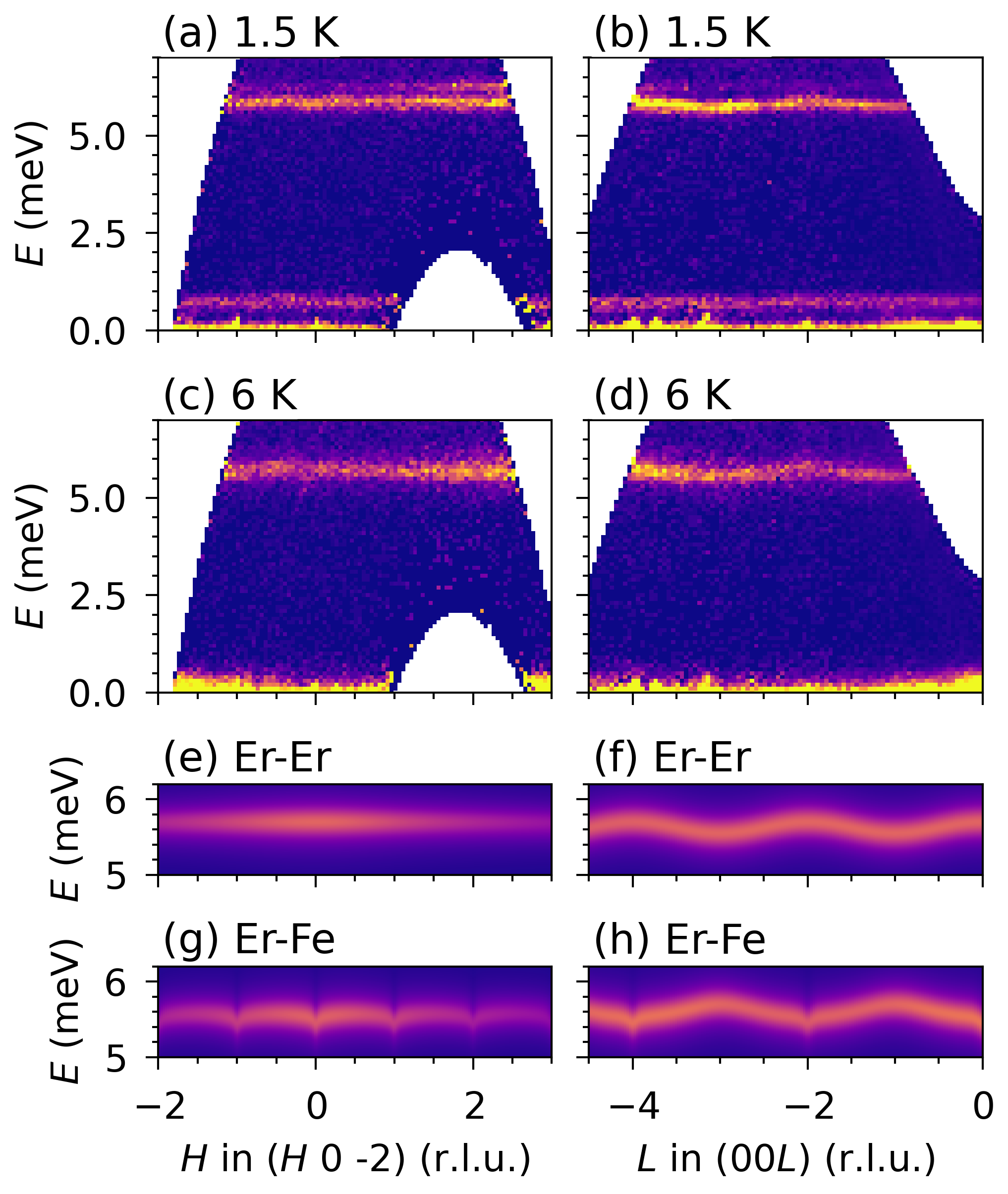}
     \caption{Dispersion and splitting of crystal electric field excitations in ErFeO$_3$ measured with $E_i$ = 9.84~meV. Excitations are mapped along the main directions in reciprocal space $(H00)$ and $(00L)$ in the left and right column panels, respectively. Spectral weight measured at (a,b) 1.5\,K and (c,d) 6\,K. Dispersive crystal field excitation modeled with (e,f) Er-Er exchange interactions and (g,h) Er-Fe exchange interactions.}
    \label{fig:CEF-dispersion}
\end{figure}

We revise the CEF scheme of Er$^{3+}$ in ErFeO$_3$ based on the CEF Hamiltonian term in Stevens formalism:
\begin{equation}
    \mathcal{H}_\mathrm{Er}^\mathrm{CEF} = \sum_{n, m} B_n^m \mathbf{O}_n^m +
    \mu_\mathrm{B} g_\mathrm{Er} \mathbf{J} \cdot \mathbf{B}_\mathrm{eff}
    \label{eq:H_Er_CEF}
\end{equation}
where $B_n^m$ are CEF parameters and $\mathbf{O}_n^m$ are Stevens operators~\cite{stevens_Onm_1952}, while the Zeeman term includes any effective fields $\mathbf{B}_\mathrm{eff}$ acting on the dipolar moment of Er ion.
The Er$^{3+}$ ion is located at the $4c$ site, where the interaction with the local environment splits the 16-fold degenerated $J=15/2$ state into 8 Kramers doublets and allows for twelve non-zero CEF parameters.
Modeling of the CEF spectra is challenging as we observed only five peaks in the measured spectra, implying the presence of accidental degeneracies or silent transitions.
Further, we observed dispersion of the lowest excitation, see \fig\ref{fig:CEF-dispersion}, which contributes to the excitation's width, and interferes with the convolution fit of the whole spectrum.

We used a set of CEF parameters constrained by the $..m$ (monoclinic, $C_s$) site-symmetry point group of Er$^{3+}$ \footnote{From the symmetry point of view, the coexistence of Fe$^{3+}$ ($\Gamma_2$, $Pbn'm'$) and Er$^{3+}$ ($\Gamma_1$, $Pbnm.1$) magnetic orderings below $T_\textrm{N}^\textrm{Er}$ leads to the monoclinic $P$~$2_1/b.1$~1~1 (non-standard setting of $P2_1/c.1$ magnetic space group) symmetry~\cite{deng_magnetic_2015} with $1$ (trivial) site-symmetry point group, which in-principle lifts all restrictions on CEF $B_l^m$ parameters. However, as magnetocrystalline coupling is weak and does not lead to the observable shifts of atomic positions, the approximation of the $Pbnm$ space group with the $..m$ site-symmetry point group of Er$^{3+}$ is justified below $T_\textrm{N}^\textrm{Er}$} and zero effective field.
We determined the initial parameters by employing the point-charge model \cite{hutchings_pointcharge_1964} in coordinate system aligned with the main crystal axes and a 3\,\AA~neighbour distance cutoff of the Er$^{3+}$ site, \ie, considering eight nearest-neighbour O$^{2-}$ ligands.
This model gives a good approximation of the observed CEF excitations energies and relative intensities at 6\,K (\fig\ref{fig:CEF-powder}c) and 120\,K (\fig\ref{fig:CEF-powder}d), as shown with the green vertical bars.
Next, we conducted least-squares refinement of the CEF parameters, by fitting the calculated spectrum convoluted with instrumental resolution to the measured one.
Some parameters gave errors larger than their value and were set to zero.
The refinement of the CEF parameters to the 6\,K data resulted in excellent agreement and yielded the spectrum showed with the red line in \fig\ref{fig:CEF-powder}c, with the parameters summarized in \tab{\ref{tab:CEF-params}}.
However, the fit results in two close-lying excitations around 6\,meV with equal intensities, see \fig\ref{fig:CEF-powder}(c), inconsistent with high-resolution measurements shown in \fig\ref{fig:CEF-dispersion}.
This originates from the dispersion of the low-lying level, which artificially broadens the excitation, and encourages the fit algorithm to fit two excitations in this region.
It remains possible that the excitation at 5.8\,meV exhibits accidental degeneracy with energy difference less than 0.1\,meV between the close lying states.

\begin{table}
\caption{
\label{tab:CEF-params}
Crystal electric field parameters of Er$^{3+}$ in ErFeO$_3$, according to the Hamiltonian term $\mathcal{H}_\mathrm{Er}^\mathrm{CEF}$ (Eq.~\ref{eq:H_Er_CEF}). The symmetry allowed parameters are labeled in the left column, the values determined from the point-charge model used as the initial values for the refinement in the middle column, and the finalized, refined parameters in the right column. The numbers in parentheses give the uncertainty in the final digits of the quoted values, for some parameters the refinement resulted in error larger than the value, and these parameters were set to zero.}
\resizebox{0.50\textwidth}{!}{  
\begin{ruledtabular}
\begin{tabular}{r r r}
Label & Point-charge (meV) & Refined (meV)         \\ \hline
$B_2^0$   & $  4.27 \cdot 10^{-2}$  & $   5.599 \, (33) \cdot 10^{-2}$ \\
$B_2^{1}$ & $ 29.13 \cdot 10^{-2}$  & $  26.04 \phantom{0} \, (12) \cdot 10^{-2}$ \\
$B_2^{2}$ & $ -1.82 \cdot 10^{-2}$  & $ 0 $ \\[0.5em] 
$B_4^0$   & $ -6.47 \cdot 10^{-4}$  & $  -4.650 \, (28) \cdot 10^{-4}$ \\
$B_4^{1}$ & $ 16.78 \cdot 10^{-4}$  & $  31.44  \, \phantom{0} (26) \cdot 10^{-4}$ \\
$B_4^{2}$ & $  9.23 \cdot 10^{-4}$  & $ 0 $ \\
$B_4^{3}$ & $-70.89 \cdot 10^{-4}$  & $ -70.58 \, (54) \cdot 10^{-4}$ \\
$B_4^{4}$ & $ -4.55 \cdot 10^{-4}$  & $ -18.47 \, (16) \cdot 10^{-4}$ \\[0.5em]
$B_6^0$   & $ -0.28 \cdot 10^{-6}$  & $  -1.350 \, ( 25) \cdot 10^{-6}$ \\
$B_6^{1}$ & $-14.88 \cdot 10^{-6}$  & $  19.85 \phantom{0} \, (29) \cdot 10^{-6}$ \\
$B_6^{2}$ & $  9.51 \cdot 10^{-6}$  & $ 0 $ \\
$B_6^{3}$ & $-39.57 \cdot 10^{-6}$  & $ -55.55 \phantom{0} \, (82) \cdot 10^{-6}$ \\
$B_6^{4}$ & $ 13.99 \cdot 10^{-6}$  & $   6.81 \phantom{0} \, (24) \cdot 10^{-6}$ \\
$B_6^{5}$ & $ 64.13 \cdot 10^{-6}$  & $ 0 $ \\
$B_6^{6}$ & $ -3.73 \cdot 10^{-6}$  & $ 0 $ \\

\end{tabular}
\end{ruledtabular}
}
\label{CEF_param}
\end{table}

To further explore the validity of the refined CEF parameters, we calculated the g-tensor of the Er$^{3+}$ ion.
It reveals strong easy axis anisotropy along the $z$ direction, that is, the $c$ axis in the $Pbnm$ setting.
This is consistent with the g-tensor determined by the optical measurements~\cite{Wood1969b}, and the $\mathcal{C}_z$ magnetic ordering of the Er-sublattice, as observed in neutron diffraction measurements \cite{gorodetsky_magnetic_1973}.
Further, we simulated the neutron scattering spectrum of ErFeO$_3$ at 120\,K, with the results shown and compared to the measured spectrum in \fig\ref{fig:CEF-powder}(d).
These results reinforce the validity of our model, however, they do not take into account the molecular field affecting erbium ions.

Finally, we observed the splitting of the ground state as shown in \fig\ref{fig:CEF-dispersion}(a,b).
Previous works report that the internal field is responsible for lifting the Kramers degeneracy of the Er ground state and splitting it by 0.4\,meV at 6\,K and 0.8\,meV at 1.5\,K \cite{zic_coupled_2021, faulhaber_optical_1967}, consistent with our observations, see \fig\ref{fig:CEF-dispersion}(a-d).
The internal field was ascribed to the Er-Fe interaction by means of dipolar and exchange interactions with estimated contributions of $1.39$~cm$^{-1}$=0.17\,meV for dipolar and $1.78$~cm$^{-1}$=0.22\,meV for exchange interactions at 6\,K \cite{faulhaber_optical_1967, Wood1969a}.
This can be incorporated to the CEF model in \eq{\ref{eq:H_Er_CEF}} by introducing an effective field along the $c$ axis direction, incorporating dipolar and exchange interactions.
The field of 0.5\,T introduces splitting of the ground state by 0.4\,meV and the first excited state of 1\,meV, as in our measurements.
Theoretical calculations \cite{faulhaber_optical_1967} suggest the 0.2\,T dipolar field from Fe sublattice, and equal contribution from Er-Fe exchange interactions, consistent with our results.

The interaction between Er- and Fe-sublattices should result in hybridization of the CEF and magnon excitations and potentially anticrossing.
We did not observe such effects in the measured spectra, see for example \fig\ref{fig:magnons_gap}, suggesting the effect is below the measurements resolution or manifests in unexplored regions of reciprocal space.
On the other hand, we have observed a clear dispersion of the 5.8\,meV CEF excitation along the $(00L)$ direction as in \fig\ref{fig:CEF-dispersion}(b,d), with no dispersion along the $(H00)$ \fig\ref{fig:CEF-dispersion}(a,c) or $(0K0)$ (not shown) directions.
The oscillation is sinusoidal, with maxima at even $L$ positions on the $(00L)$ line and bandwidth of 0.25\,meV, suggesting antiferromagnetic coupling between ions stacked along the $c$-axis.

To elucidate the observed dispersive nature of the Er$^{3+}$ CEF excitation, we utilized the linear spin wave theory including the erbium sublattice.
We considered effective spin operators $\mathcal{S}$ for erbium ions that accommodate the spin number J=15/2 and couple with strength $\mathcal{J}_n$, where $n$ indexes the bond.
We take only quadratic anisotropy terms, which result in easy axis roughly along the $c$-axis $\mathcal{K}_z<0$ \footnote{The $B_2^0 O_2^0+B_2^1 O_2^1$ anisotropy is equivalent to $B_2^0 (3J_z^2-1)+\frac{1}{2}B_2^1(J_xJ_z+J_zJ_x)$. For $B_n^m$ values from \tab{\ref{tab:CEF-params}} this gives tilted anisotropy ellipsoid, with the main (easy) axis vector tilted 30 degrees away from the $z$-axis toward the $x$ axis.}.
The minimal model that describes the dispersion of the lowest CEF level is described by the hamiltonian
\begin{equation}
    \label{eq:H_Er-Er_Ex}
    \mathcal{H}_\mathrm{Er} = - \mathcal{J}_c \sum_{i,j \in \mathrm{NN}_c} \mathbf{\mathcal{S}}_i \cdot \mathbf{\mathcal{S}}_j + \sum_{i} \mathcal{K}_z (\mathcal{S}_i^z)^2.
\end{equation}
It contains the anisotropy term with refined value $\mathcal{K}_z$=$-0.3799 (16)$ meV, and only a single coupling term between nearest Er ions forming the chain along the $c$-axis direction, see \fig\ref{fig:Interactions} (b), with coupling strength  $\mathcal{J}_c$=$0.00489 (74)$ meV.
These parameters were obtained by refining the hamiltonian parameters to the dispersion relations of the lowest CEF excitation measured at 6\,K, see \fig\ref{fig:CEF-dispersion}(c,d), with the modeled spectral weight shown in \fig\ref{fig:CEF-dispersion}(e,f).
The most important result is that the single antiferromagnetic coupling between intrachain Er ions fully reproduces the characteristics of the measured spectra: dispersive character for the $(00L)$ direction, and no significant dispersion for the in-plane directions.

Following previous optical studies \cite{faulhaber_optical_1967, Wood1969a, kim_observation_2025} we also included (i) the dipolar interactions and (ii) nearest neighbour Er-Fe exchange interaction.
Dipolar interactions introduce very weak dispersion of excitations across all propagation vectors and bandwidth of 0.05\,meV \footnote{The dipolar interactions involve all ions and take magnetic moments as parameters, that is 5\,$\mu_\mathrm{B}$ for iron and 9\,$\mu_\mathrm{B}$ for erbium.}.
Exchange interactions between Er and Fe sublattices introduce dispersion along the c-axis direction, with the minima at even L positions, regardless of the exchange interaction sign, as shown in \fig\ref{fig:CEF-dispersion}(g,h).
This is due to the G-type ordering of the Fe sublattice: Er-Fe bonds related by the $m_{001}$ mirror plane include Fe ions with antiparallel spins, thereby resulting in fully compensated exchange field.
In consequence, the Er-Fe exchange enters the dispersion relation in terms of absolute value that disregards the sign of the interaction: $\omega \propto \sqrt{A^2 - |J_\mathrm{Er-Fe}|^2}$, where $A$ contains single-ion anisotropy.

In conclusion of this section, the coupling between Er ions forming the chain running along the $c$ axis fully reproduces the characteristics of the measured spectra.
However, there are limitations to our model, as we are not able to reproduce the splitting of the ground state Kramers doublet of 0.8\,meV observed at 1.5\,K.
The splitting was attributed to the Er-Fe coupling by means of exchange \cite{Wood1969a} and dipolar interactions \cite{faulhaber_optical_1967}, which do not comply with our observations.
In addition, the $\mathcal{J}_c$ coupling is antiferromagnetic, while diffraction measurements showed a ferromagnetic arrangement of the intrachain ions \cite{gorodetsky_magnetic_1973}.
These qualities would not be conflicting in the presence of a stabilization mechanism arising from higher energy scales.
Our measurements provide a comprehensive overview of the relevant energy scales, summarized in \tab\ref{tab:hamiltonian}, none of which stabilize the ferromagnetic arrangement.
Within our model the antiferromagnetic arrangement of Er chains is energetically favourable, contrary to the diffraction measurements.


\begin{table*}
\caption{
\label{tab:hamiltonian}
An overview of the Hamiltonian terms in eq.~(\ref{eq:hamiltonian_general}) and their corresponding energy scales.
A symbol, a defining equation and a short description are provided for each term in the 'Term', 'Equation' and 'Description' columns, respectively.
The 'Spectrum' column denotes the energy scale at which the given term influences the excitation spectrum.
The 'Parameters' column includes refined values of the interaction parameters,
where the Fe-Er dipolar interaction, $\mathcal{H}_\mathrm{Fe-Er}^\mathrm{Dip}$, is self-consistent, i.e. it does not contain any refineable parameter.
The 'Experimental evidence' column lists the physical phenomena that require the given Hamiltonian term for modelling.
Those phenomena are gathered according to the experimental techniques: inelastic neutron scattering (INS), neutron diffraction (ND), magnetometry (MAGN), and optical studies (OPT) with the corresponding literature references.
The corresponding figures are given for all evidence marked 'INS'.
The 'FM' and 'AFM' stand for 'ferromagnetic' and 'antiferromagnetic', respectively.}
\resizebox{1.00\textwidth}{!}{%
\begin{ruledtabular}
\begin{tabular}{llllll}

Term                                                       & Eq.                                        & Spectrum                                          & 
Parameters                                                 & Description                                & Experimental evidence                             \\ \hline

\multirow{2}{*}{$\mathcal{H}_\mathrm{Fe-Fe}^\mathrm{Ex}$}  & \multirow{2}{*}{(\ref{eq:H_Fe-Fe_Ex})}     & \multirow{2}{*}{$\sim$65\,meV}                    &
$J_1=5.09\,(46)$\,meV                                          & Exchange interaction                       & Fe spin wave dispersion 
                                                                                                (Figs. \ref{fig:magnons_120K}, \ref{fig:magnons_Ecut}): INS \\

                                                           &                                            &                                                   &
$J_2=0.31(18)$\,meV                                          & within  Fe sublattice                      & Fe G-type magnetic ordering: ND
                                                                                    \cite{koehler_neutron_1960,gorodetsky_magnetic_1973,deng_magnetic_2015} \\[0.5em]

\multirow{2}{*}{$\mathcal{H}_\mathrm{Er}^\mathrm{CEF}$}    & \multirow{2}{*}{(\ref{eq:H_Er_CEF})}       & \multirow{2}{*}{$\sim$36\,meV}                     &
$B_n^m$ see Table~\ref{tab:CEF-params}                     & Er crystal electric field  & Er CEF levels presence 
                                                                               (Fig. \ref{fig:magnons_gap}): INS \cite{zic_coupled_2021, obrien_giant_2023} \\
                                                                            
                                                           &                                            &                                                   &
$\mathcal{K}_c = -0.3799 \, (16)$\,meV                            & Er effective anisotropy                    & Fe spin reorientation: ND
                                                           OPT \cite{faulhaber_optical_1967, Yamaguchi_2013_Terahertz, Mikhaylovskiy_2017_Selective, kim_observation_2025} \\[0.5em]
                                                           
\multirow{2}{*}{$\mathcal{H}_\mathrm{Fe}^\mathrm{SIA}$}    & \multirow{2}{*}{(\ref{eq:H_Fe_SIA})}       & \multirow{2}{*}{$\sim$9\,meV}                     &
$K_a\textbackslash K_c$                                    & \multirow{2}{*}{Fe single-ion anisotropy}  & Fe spin wave energy gap 
                                                                               (Fig. \ref{fig:magnons_gap}): INS \cite{zic_coupled_2021, obrien_giant_2023} \\
                                                                            
                                                           &                                            &                                                   &
$= -0.117\,(35)$\,meV                                             &                                            & Fe spin reorientation: ND
                                                                                    \cite{koehler_neutron_1960,gorodetsky_magnetic_1973,deng_magnetic_2015} \\ [0.5em]

\multirow{3}{*}{$\mathcal{H}^\mathrm{Dip}$} & \multirow{3}{*}{(\ref{eq:H_Dip})}    & \multirow{3}{*}{$\sim$0.5\,meV}                   &
\multirow{3}{*}{$B_\mathrm{eff} = 1$\,T}     & Er effective field from & Er CEF splitting
                                             (Fig. \ref{fig:CEF-dispersion}): INS \cite{zic_coupled_2021, obrien_giant_2023}                                \\

                                                           &                                            &                                                   &
                                                           &  dipolar and exchange & and OPT \cite{faulhaber_optical_1967, Yamaguchi_2013_Terahertz, Mikhaylovskiy_2017_Selective, kim_observation_2025} \\
                                                                                    
                                                           &                                            &                                                   &
                                                           &  interactions   & Fe weak AFM canting: ND
                                                                                    \cite{koehler_neutron_1960,gorodetsky_magnetic_1973,deng_magnetic_2015} \\ [0.5em]

\multirow{2}{*}{$\mathcal{H}_\mathrm{Fe-Fe}^\mathrm{DMI}$} & \multirow{2}{*}{(\ref{eq:H_Fe-Fe_DMI})}    & \multirow{2}{*}{$\sim$0.5\,meV}                         &
\multirow{2}{*}{$D_y=0.3$\,meV}                        & Antisymmetric exchange                     & Fe spin wave degeneracy lifting (not observed)    \\
                                                                                                     
                                                           &                                            &                                                   &
                                                           & within Fe sublattice                       & Fe weak FM canting: 
                                                          MAGN \cite{bozorth_magnetization_1958, bazaliy_reorientation_2004, fita_temperature-driven_2022}  \\ [0.5em]
                                                                                    
\multirow{2}{*}{$\mathcal{H}_\mathrm{Er-Er}^\mathrm{Ex}$}  & \multirow{2}{*}{(\ref{eq:H_Er-Er_Ex})}     & \multirow{2}{*}{$\sim$0.1\,meV}                   &
\multirow{2}{*}{$\mathcal{J}_c=0.00489 \, (74)$\,meV}              & Exchange interaction                       & Er CEF levels dispersion
                                                                                               (Fig. \ref{fig:CEF-dispersion}): INS \cite{zic_coupled_2021} \\
                                                                                                                                    
                                                           &                                            &                                                   &
                                                           & within Er sublattice                       & Er C-type magnetic ordering: ND
                                                                                    \cite{koehler_neutron_1960,gorodetsky_magnetic_1973,deng_magnetic_2015} \\
\end{tabular}
\end{ruledtabular}
}%
\end{table*}

\section{Discussion}
\label{sect:GS}

Our inelastic thermal neutron scattering data give direct access to the high- and intermediate-energy scales of the magnetic interactions in ErFeO$_3$.
At the lowest energy scale, the dipolar and Dzyaloshinskii–Moriya interactions govern the fine structure of magnetic arrangement and the interplay between Er and Fe sublattices.
They manifest themselves in the static spin correlations, accessed by magnetization and diffraction measurements.

In particular, magnetization measurements \cite{bazaliy_reorientation_2004, fita_temperature-driven_2022} have shown the presence of a weak ferromagnetic moment (WFM) of $F\approx 5\cdot10^{-2} \mu_\mathrm{B}$ per formula unit ErFeO$_3$, which changes its orientation from $F_z$ above $T^\mathrm{Fe}_\mathrm{SR}$ to $F_x$ below $T_\mathrm{SR}$.
This weak FM component is common in orthoferrites and arises from the DMI between the Fe$^{3+}$ ions \cite{podlesnyak_RFO_2021, hahn_inelastic_2014, nikitin_decoupled_2018}.
To simulate the tilting of antiferromagnetic component of Fe$^{3+}$ moment within the $ac$-plane, we introduce to the Hamiltonian the DMI between the Fe$^{3+}$ ions within the $ab$-plane, with a single $D_y$ component:
\begin{equation}
    \label{eq:H_Fe-Fe_DMI}
    \mathcal{H}_\mathrm{Fe-Fe}^\mathrm{DMI} = \sum_{ij \in \mathrm{NN}_{ab} } D_\mathrm{y} \, \hat{y} \cdot \left( \mathbf{S}_i \times \mathbf{S}_j \right),
\end{equation}
where the summation takes nearest-neighbours in the $ab$-plane, $\mathbf{S}_i$ and $\mathbf{S}_j$ are Fe spins, and $\hat{y}$ denotes the unit vector along the $b$-axis.
The value of $D_y=0.15$\,meV accurately reproduces the WFM observed in magnetization measurements~\cite{bazaliy_reorientation_2004, fita_temperature-driven_2022}, and corresponds to the canting of Fe$^{3+}$ moments of around 0.5\degree, very similar to the case of YFeO$_3$ \cite{hahn_inelastic_2014}, YbFeO$_3$ \cite{nikitin_decoupled_2018}, and NdFeO$_3$ \cite{kumar_exchange_2025}.
Important to note, this single DMI component gives rise to tilting above and below the spin reorientation transition.

The spin reorientation transition in orthoferrites is a well studied phenomenon, rooted in the temperature dependent magnetocrystalline anisotropies arising from CEF \cite{white_review_1969}.
At high temperatures, where CEF levels above the ground state of Er are occupied, the Er magnetic moment is roughly isotropic and the Fe ion will follow its own single-ion anisotropy, that is order the moments along the $a$-axis.
At low temperature Er 4f electrons occupy only the highly anisotropic ground state, introducing effective anisotropy on Fe$^{3+}$ ions.
As shown here, the CEF causes Er ions to align its magnetic moment along the $c$ axis, and the Fe spins follow.
The reorientation thus occurs at temperatures relating to the first excited state $k_\mathrm{B}T$=5.8\,meV giving $T=$70\,K in agreement with $T^\mathrm{Fe}_\mathrm{SR}\approx$\,88 -- 98\,K, but more importantly, explains its continuous character.
It also explains why orthoferrites with non-magnetic rare-earth element, LaFeO$_3$, LuFeO$_3$, YFeO$_3$, do not undergo spin reorientation \cite{white_review_1969}.
Here, we used an effective model for description of Fe anisotropies across the spin reorientation, as detailed in \sect\ref{sect:spin waves}.

Neutron diffraction measurements report an additional antiferromagnetic $C_y$ component of the Fe sublattice ordering below the $T_\mathrm{N}^\mathrm{Er}$ \cite{gorodetsky_magnetic_1973, deng_magnetic_2015}.
As the $C_y$ component is concomitant with the development of the long-range order of the Er sublattice, we suggest that the interlattice coupling influences the Fe sublattice order by means of dipolar interactions.
In the Hamiltonian we include the term
\begin{equation}
    \mathcal{H}^\mathrm{DIP} = 
    - \sum_{\langle i,j \rangle}
    \frac{\mu_0}{4\pi}
    \left[ 
    \frac{3(\mathbf{m}_i \cdot \mathbf{R_{ij}})(\mathbf{m}_j \cdot \mathbf{R_{ij}})}{|\mathbf{R_{ij}}|^5} - \frac{\mathbf{m}_i \cdot \mathbf{m}_j}{|\mathbf{R_{ij}}|^3}
    \right],
    \label{eq:H_Dip}
\end{equation}
where $\big<i,j\big>$ enumerates pairs of interacting magnetic moments $\mathbf{m}_i$ connected with the bond vector $\mathbf{R}_{ij}$.
This couples all sublattices, that is Fe-Fe, Er-Fe and Er-Er.
We use the Ewald summation method as implemented in \textsc{Sunny}~\cite{sunny_2025}.
As a starting point of the calculation, we take the collinear arrangement of magnetic moments with $G_z$ for Fe and $\mathcal{C}_z$ for Er sublattices.
Minimizing the energy of the system results in the $G_y\boldsymbol{G_z}$ arrangement for the Fe sublattice, and $\mathcal{C}_z$ for Er sublattice.
Thus, the arrangement type of the Fe sublattice is in agreement with an older report \cite{gorodetsky_magnetic_1973}, but contrary to the recent one \cite{deng_magnetic_2015}.
Both $G_y$ and $\mathcal{C}_z$ follow the same irrep, $\Gamma_1$, consistent with the idea that Fe$^{3+}$ $G_y$ component is induced by the interaction with the ordered Er$^{3+}$ sublattice.
The tilting of the moments from the $c$-axis reported at 1.5\,K is 48\degree \cite{gorodetsky_magnetic_1973}, while our calculations yield 8\degree.

We summarize our model in \tab\ref{tab:hamiltonian} based on the spectroscopic signatures.
It reveals a hierarchy of the interaction scales and their characteristic energy, typical for orthoferrites.
The system is dominated by strong, antiferromagnetic exchange interaction between iron spins, giving rise to the G-type magnetic order with high transition temperature.
The single-ion anisotropies of either Er or Fe ions are in the intermediate energy scale.
The iron spins exhibit easy axis anisotropy affected by temperature-dependent magnetocrystalline anisotropies, originating from change in the occupation of CEF levels of Er ions.
At the lowest range lie Er-Er exchange interactions, dipolar interactions and DMI.
The hierarchy of interaction scales is also reflected in the characteristic temperatures of the system: Fe-Fe exchange sets at T$_\mathrm{N}^\mathrm{Fe}$ = 620 K, magnetocrystalline anisotropies result in spin reorientation at $T^\mathrm{Fe}_\mathrm{SR}\approx$\,88 -- 98\,K, and the Er-related interactions setting in below $T_\mathrm{N}^\mathrm{Er}$=4.1\,K.

Throughout this work, we have utilized linear spin-wave theory to model dispersive excitations, as well as the Stevens formalism to model crystal field excitations.
In order to explain the dispersion of the first excited state and the splitting of the ground state an approach incorporating two methods is required.
It is apparent that constraining the physics of the Er sublattice to the 1/S approximation as per LSWT does not allow us to fully describe the spectral signatures in the system and may account for some inconsistencies between our model and the experimental observations.
On the other hand, incorporating the full manifold of J=15/2 states for Er and S=5/2 for Fe and coupling them is beyond current methods.
Orthoferrites thus host a rich interplay of multipolar and spin-wave physics, making them a complex and fascinating class of magnetic materials.

\section{Summary and Conclusions}

This work presents a comprehensive investigation of the magnetic excitation spectrum and establishes a hierarchy of magnetic interactions in ErFeO$_3$, considering both Fe$^{3+}$ and Er$^{3+}$ sublattices.
Our results are consistent with previous neutron diffraction \cite{gorodetsky_magnetic_1973, deng_magnetic_2015} and low energy spectroscopy \cite{faulhaber_optical_1967, zic_coupled_2021} studies and allow for deeper insight into the magnetic interactions, owing to momentum-resolved spectroscopy.
In particular, we report the thorough Fe$^{3+}$ spin wave dispersion over the entire Brillouin zone, a revision of the Er$^{3+}$ CEF energy levels, and dispersive character of the CEF levels.
These findings and incorporation of previous reports enable us to construct a Hamiltonian that consistently accounts for both the magnetic ground state and the excitation spectrum of ErFeO$_3$ over all energy scales in the system.

Our results reveal a marked contrast between the energy scales governing the two magnetic subsystems. 
In the Fe$^{3+}$ sublattice, the exchange interactions dominate over magnetocrystalline anisotropy, whereas in the Er$^{3+}$ sublattice the crystal electric field constitutes the leading energy scale, greatly exceeding the Er–Er exchange interaction.
The coupling between the Fe and Er subsystems is comparatively weak and arises from a combination of magnetic dipole – dipole and exchange interactions, yet it plays a crucial role in CEF splitting and in stabilizing the low-temperature magnetic ground state.

\section*{Acknowledgments}

The sample was grown within a project funded by DFG (SA-3688/1-1) and RFBR (19-52-12047).
We acknowledge the provision of beamtime on MERLIN at ISIS spallation source through proposal no. RB2510189. 
DRB acknowledges GNeuS funding from the European Union's Horizon 2020 research and innovation program under the Marie Skłodowska-Curie grant agreement No. 101034266.

\section*{Data Availability Statement}

The neutron spectroscopy data that support the findings of this article are openly available \cite{Data_Merlin_ISIS_2025}, and the rest of the data are available upon reasonable request from the authors.

\bibliography{EFO_Bibtex}  

\end{document}